# Thickness-dependence of Linear and Nonlinear Optical Properties of Multilayer 3R-MoS$_2$

*Fatemeh Abtahi[1], Alen Shaji[1], Gia Quyet Ngo[1,2], Benjamin Laudert[1], Hossein Esfandiar[2], Sebastian W. Schmitt[1,2], and Falk Eilenberger[1,2,3]*

[1] *Institute of Applied Physics, Abbe Center of Photonics, Friedrich Schiller University Jena, Albert-Einstein-Str. 15, 07745 Jena, Germany*

[2] *Fraunhofer Institute for Applied Optics and Precision Engineering IOF*

*Albert-Einstein-Str. 7, 07745 Jena, Germany*

[3] *Max Planck School of Photonics, Albert-Einstein-Str. 15, 07745 Jena, Germany*

**Abstract:** 3R-MoS$_2$, a MoS$_2$ polytype with broken inversion symmetry, enables unique light-matter interactions and is promising for linear and nonlinear integrated photonics beyond the monolayer limit. Yet, systematic studies of its thickness-dependent reflectivity and its impact on harmonic generation are still lacking. Here, we introduce a non-destructive optical method to determine the thickness of 3R-MoS$_2$ flakes from reflectivity measurements, offering AFM-like precision with a mean bias of less than 2 nm, while being much faster and applicable to non-solid substrates such as PDMS, in the 3-200 nm range. Nonlinear characterization further reveals distinct thickness-dependent maxima in second- and third-harmonic generation (SHG / THG), with the first clear peak at ~200 nm. These maxima arise from Fabry-Pérot-type phase matching conditions mediated by the film thickness and can further be shaped by absorption. This work thus provides both a practical thickness metrology and new insights for exploiting thickness-dependent 3R-MoS$_2$ nonlinearities in scalable photonic technologies.

## 1. Introduction

Van der Waals transition metal dichalcogenides (TMDs) such as molybdenum disulfide (MoS$_2$)[1,2] are noted for their strong light-matter interaction,[3–5] and tunable electronic structure[6,7]. While the 2H polytype of MoS$_2$ has been studied extensively[8–11], the rhombohedral 3R phase, the second most stable stacking sequence, offers qualitatively different behavior. Its unidirectional layer arrangement breaks inversion symmetry even in bulk[12], localizes excitonic wavefunctions, extends exciton lifetimes[13,14], and results in notable piezoelectric and nonlinear optical responses[2,12,15–17]. In particular, the piezoelectric constant of 3R-MoS$_2$ peaks at five layers despite diminishing in multilayer 2H-MoS$_2$ [18], highlighting how stacking order and interlayer coupling can radically reshape functional properties.

When light illuminates a TMD, it may be reflected, transmitted or absorbed; each channel depends sensitively on layer number, stacking sequence and defects. In 3R-MoS$_2$, the absorption spectrum is dominated by two excitonic resonances at 1.9 and 2.1 eV[19] that remain unchanged in energy regardless of layer count, and indicates that interlayer interactions effectively isolate excitons[2,19,20]. Its bandgap evolves from direct in the monolayer to indirect in the bulk, with increasing thickness[15,21,22]. Accurate determination of flake thickness is therefore a requirement for correlating these optical signatures with structure. Non-destructive methods such as ellipsometry[23,24], Raman spectroscopy[25,26] and photoluminescence mapping[27] are commonly used, and optical reflectivity measurement, especially when merged with computational modeling, has emerged as a complementary approach to estimate thickness with sub-nanometer precision[28]. Yet, practical implementation is challenged by substrate interference (e.g. on PDMS) and limited availability of high-quality 3R-MoS$_2$ flakes. The main preparation methods for this material family rely on exfoliation and subsequent transfer printing by PDMS. However, direct thickness determination of flakes on PDMS remains challenging, as AFM cannot be applied in this configuration easily. A non-destructive optical approach would therefore be highly desirable.

Beyond linear optics, 3R-MoS$_2$'s broken inversion symmetry makes it a powerful platform for second- and third-order nonlinear processes[17,29]. Stable out-of-plane valley and spin polarization have already been demonstrated[16]. Second-harmonic generation (SHG) cannot be supported by centrosymmetric materials and it only emerges where inversion symmetry is broken, for example at monolayers, surfaces, and interfaces[30]. Studies of SHG report higher efficiency in thick 3R-MoS$_2$ than in the 2H phase[31,32]. However, there exists no comprehensive study of third-harmonic generation (THG) versus flake thickness in the 3R phase.

THG is a third order nonlinear optical process where three photons of the same frequency combine to produce a single photon at triple the frequency, $3\omega$ ($\lambda_{3\omega} = \lambda_\omega/3$)[33]. Unlike SHG, THG is symmetry-independent, yielding a uniform polarization pattern even in centrosymmetric and even-layered TMDs[34]. TMDs exhibit strong third-order susceptibility $\chi^{(3)}$ and experiments have shown that their THG response is highly sensitive to thickness, excitation wavelength, and the material's absorption resonances.

In 2H-MoS$_2$ monolayer, THG intensity is further enhanced when the $3\omega$ emission overlaps with an exciton[35,36], boosting the conversion efficiency by up to an order of magnitude. To date, however, studies of layer dependence THG in TMDs have been confined to just a few layers[34] and have not addressed the 3R-phase of MoS$_2$. Extending the characterization of THG across a broad thickness range in this less explored polytype is therefore essential, both to complete our understanding of nonlinear responses in TMDs and to pave the path toward potential applications in all-optical switching[37] and quantum optics[29], where entangled photon sources are highly desirable.

In this work, we address these gaps by developing a robust, non-destructive method for 3R-$MoS_2$ thickness characterization, integrating calibrated optical reflectivity, numerical modeling and AFM validation on flakes exfoliated onto PDMS and transferred to $SiO_2$, and by performing the first systematic investigation of THG intensity as a function of thickness for 3R-$MoS_2$ flakes. Our results establish robust reference for non-invasive thickness determination and demonstrate the scaling relationship governing nonlinearity in 3R-$MoS_2$, opening the way for its use in advanced nonlinear photonic[17], quantum technologies[29], micro-resonators[4], and all-in-fiber integrated photonics[38–42].

## 2. Results and discussion

Details of the sample preparation and experimental setups for both linear and nonlinear characterizations are provided in the 'material and methods' section.

### 2.1. 3R-$MoS_2$ thickness calculation through reflectivity measurement

**Figure 1**a demonstrates one of the exfoliated 3R-$MoS_2$ flakes on PDMS with marked region 1 (blue) and 2 (red). **Figure 1**b and c show the measured reflection spectra (experimental data in blue color) of region 1 and 2 on 3R-$MoS_2$ flake exfoliated on a PDMS substrate for the wavelength range of 600 to 1000 nm. The details of the reflectivity measurement are explained in 'material and methods' section. The spectra were recorded and subsequently normalized to a protected silver mirror reference to yield absolute reflectivity data. The reproducibility of the measurement has been checked with a test sample. To interpret the measured spectra, a theoretical model for including reflection, transmission, and absorption was developed (Supporting Information (S1.1)). It is governed by 3R-$MoS_2$'s refractive index taken from from Xu et al. [4] and the incident wavelength, with the thickness as a free model parameter.

The numerical model was fitted (numerical model in orange color) to this experimental data to estimate the thickness of each region. The thickness estimated for region 1 and 2 of the flake on PDMS with reflective measurement is 185 nm and 200 nm. We then successfully transferred the same flake from PDMS to an $SiO_2$ substrate. **Figure 1**d shows the microscopic image of the flake on $SiO_2$ substrate. To further probe the reflectivity-based thickness determination, a similar measurement has been done on the same regions of the flake on the $SiO_2$ substrate. **Figure 1**e and f display the measured reflection spectrum of 3R-$MoS_2$ on $SiO_2$ for the same wavelength range of 600 nm to 1000 nm and the model adapted to account for the $SiO_2$ substrate properties and fitted to the experimental data, which results in consistent thickness estimates of 185 nm and 200 nm, respectively.

Atomic force microscopy (AFM) was used to validate the thickness estimation based on the optical reflection spectrum. **Figure 1**g highlights the AFM profile of region 1 and 2 on the flake on the SiO$_2$ substrate indicates thicknesses of 185 ±2.4 nm and 200 ±2.2 nm, respectively.

To validate the proposed technique for a large range of thickness values, this procedure was then repeated on a series of 3R-MoS$_2$ flakes with varying thicknesses from 3 nm to 200 nm. The results of this shown in **Figure 1**h demonstrate the consistency of the thickness measured of the method for a large range of thicknesses and to create enough data for statistical analysis. The Bland-Altman analysis showed a mean bias of Δ= -1.67 nm between reflectometry and AFM. The limits of agreement (LoA) were -8.40 to 5.07 nm, indicating that for ~95% of measurements the difference between the two methods falls within this narrow range. This demonstrates good agreement, with only a small systematic offset of reflectometry relative to AFM. Corresponding Bland-Altman plot is provided in Supporting Information (S1.3). This highlights the potential of the method to directly determine the 3R-MoS$_2$ flake thickness on a PDMS substrate with nanometer precision. It eliminates the need to transfer samples onto a rigid substrate for AFM and therefore facilitates the direct transfer of samples to the desired functional substrates, such as photonic integrated circuits.

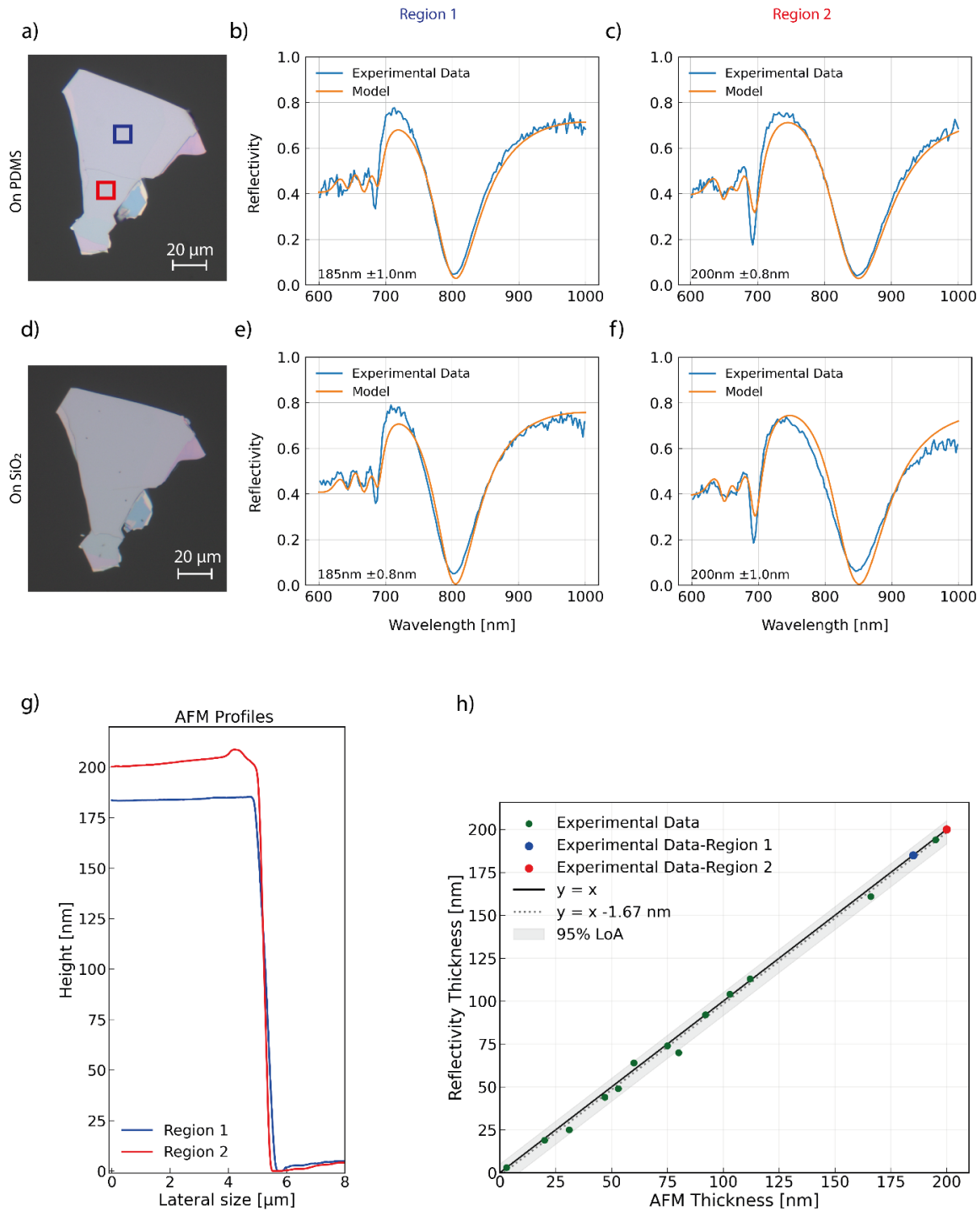

**Figure 1 3R-MoS$_2$ flake thickness determined by reflection measurements and AFM**. a) microscopic image of exfoliated 3R-MoS$_2$ on PDMS with marked regions of reflectivity and AFM measurement. Region 1 (blue) and Region 2 (red). b) measured reflection spectra of 3R-MoS$_2$ with a visible light in region 1 and c) region 2 and numerically calculated thickness based on the best fitted model on PDMS. d) Microscopic image of transferred 3R-MoS$_2$ on SiO$_2$. After successful transfer, the reflectivity is measured again on e) region 1 and f) region 2 and the fitted numerical model shows a good agreement with the reflectivity results with PDMS as substrate. g) Height profile of the region 1 and 2 using AFM measurement. h) Thickness measured on PDMS using reflectivity measurement vs. using AFM measurement for the range of thickness from 3 to 200 nm shows a linear dependency.

## 2.2. Thickness dependence of Second Harmonic Generation of 3R-MoS$_2$

Having quantified the 3R-MoS$_2$ flake thicknesses, we now want to probe their thickness dependence nonlinear optical responses. The second harmonic generation (SHG) and third harmonic generation (THG) measurements were carried out using a custom-built nonlinear microscope setup shown in **Figure 2**a, femtosecond pulses at 1600 nm were focused onto the samples and the SHG and THG signals were collected in transmission through appropriate filters for each nonlinear process. **Figure 2**b shows the SHG spectrums for different input power on the 3R-MoS$_2$ flake on SiO$_2$ substrate. The normalized SHG spectra are located around 800 nm centered at half of the fundamental wavelength (1600nm). In **Figure 2**c we show the quadratic dependence of the SHG intensity to the excitation power. The results are presented in a log-log plot, resulting in a slope of 1.98, close to the expected square power. This gives evidence for the SHG measured from our 3R-MoS$_2$ flakes. In the next step the SHG thickness dependence of the 3R-MoS$_2$ flakes was investigated. To ensure a valid comparison of the SHG signals over all different 3R-MoS$_2$ flakes, which are highly sensitive to the crystallographic orientation of the sample, we first determined the armchair crystalline axis of the flake via polarization-dependent measurements (**Figure 2**d), finding that the armchair direction for the selected flake is $\theta = 7.15°$. In this measurement, by using a fixed linear polarizer and half-wave plate to control the input polarization and a linear polarizer used as an analyzer before our detector, the input and output polarization were rotating in parallel to each other. We then aligned the incident polarization parallel to this direction before acquiring the SHG map. This procedure guarantees that all SHG data were collected under identical conditions and that the signal is maximized by orienting the input polarization along the armchair axis of the flake. **Figure 2**e presents the SHG intensity map of the flake on a logarithmic scale, highlighting the dependence of the nonlinear response on thickness of the flake. We clearly observe the higher SHG signals in region 2 (thickness ~200nm) compared to the region 1 (thickness ~ 185nm).

**Figure 2**f shows the thickness dependence of SHG from both experiment and modeling. The exfoliated 3R-MoS$_2$ flakes we used for SHG measurements have thicknesses between 20 and 620 nm. In this experiment, the excitation wavelength was fixed at 1600 nm, which generated the SH at 800 nm. The second harmonic intensity modulates as a function of the thickness (Supporting Information (S3)), with distinct resonances induced the interplay of Fabry-Pérot resonances and phase matching. The intensity of those resonances increases because the propagation loss at the FW and SH are almost negligible at their respective wavelengths.

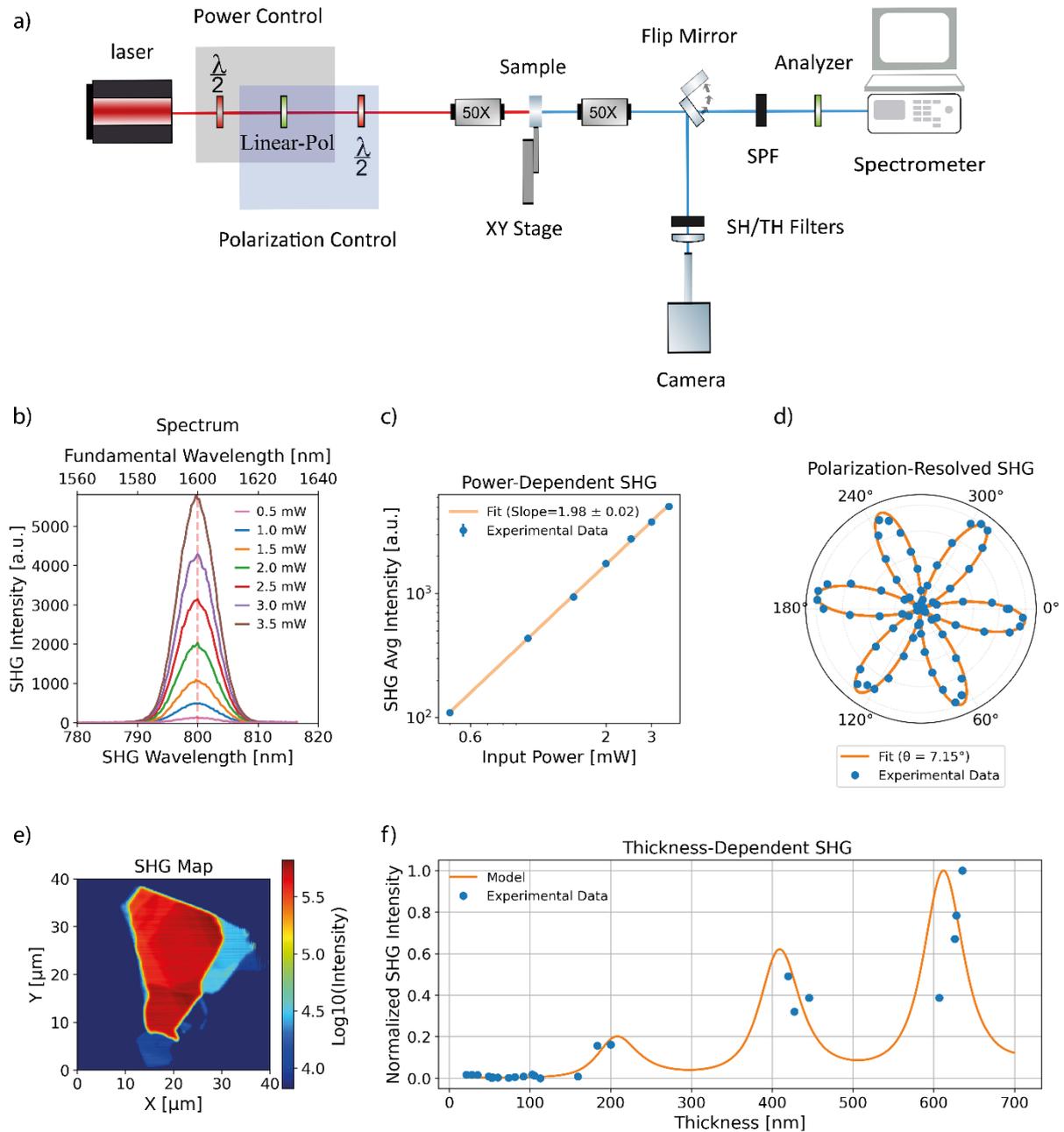

**Figure 2 Second Harmonic Generation (SHG) of 3R-MoS$_2$** a) sketch of the nonlinear microscope setup for doing second and third harmonic generation. b) SHG spectra of 1600 nm excitation wavelength for different input power. Orange dashed line shows the expected center of the spectrum at 800 nm. c) Power dependence of SHG intensity, plotted on a log-log scale and fitted with a linear function (solid orange line). d) polarization resolved SHG to find the armchair direction of the 3R-MoS$_2$ flake (θ =7.15°). e) SHG microscopic measurement when the input polarization and output analyzer are parallel to the armchair direction of the 3R-MoS$_2$. f) Thickness dependent SHG of the 3R-MoS$_2$ with excitation at a fundamental wavelength of 1600 nm. The orange line shows the theoretical model for the SHG efficiency.

## 2.3. Thickness dependence of the Third Harmonic Generation of 3R-MoS$_2$

**Figure 3**a presents the Third Harmonic Generation (THG) spectra measured at different input powers for the 3R-MoS$_2$ flake on a SiO$_2$ substrate. The THG peak appears at 533 nm, and the spectra are centered at one third of the fundamental wavelength (1600 nm). **Figure 3**b shows the cubic dependence of THG intensity on excitation power in a log–log scatter plot, with a linear fit yielding a slope of 3.12, in close agreement with the cubic behavior expected for the plot in a third-order nonlinear optical process. This power dependence confirms THG generation.

**Figure 3**c shows that the THG response is independent of the flake's crystallographic orientation, with negligible variation as the input polarization is rotated. Nevertheless, all subsequent measurements were performed with the excitation polarization aligned along the armchair direction, to maintain consistency with the SHG experiments.

**Figure 3**d presents a logarithmic-scale THG intensity map of the flake, clearly revealing the dependence of the nonlinear response on thickness. While the thickness dependence of SHG in 3R-$MoS_2$ is well established, here we focus on the thickness dependence THG and compare its behaviour to SHG in this material. In the map of the studied 3R-$MoS_2$ flake, we observe stronger THG signals in region 2 (thickness ≈200 nm) as compared to region 1 (thickness ≈185 nm). We then systematically investigate the influence of thickness by measuring the THG signal across several flakes of different thicknesses, in analogy to our SHG analysis.

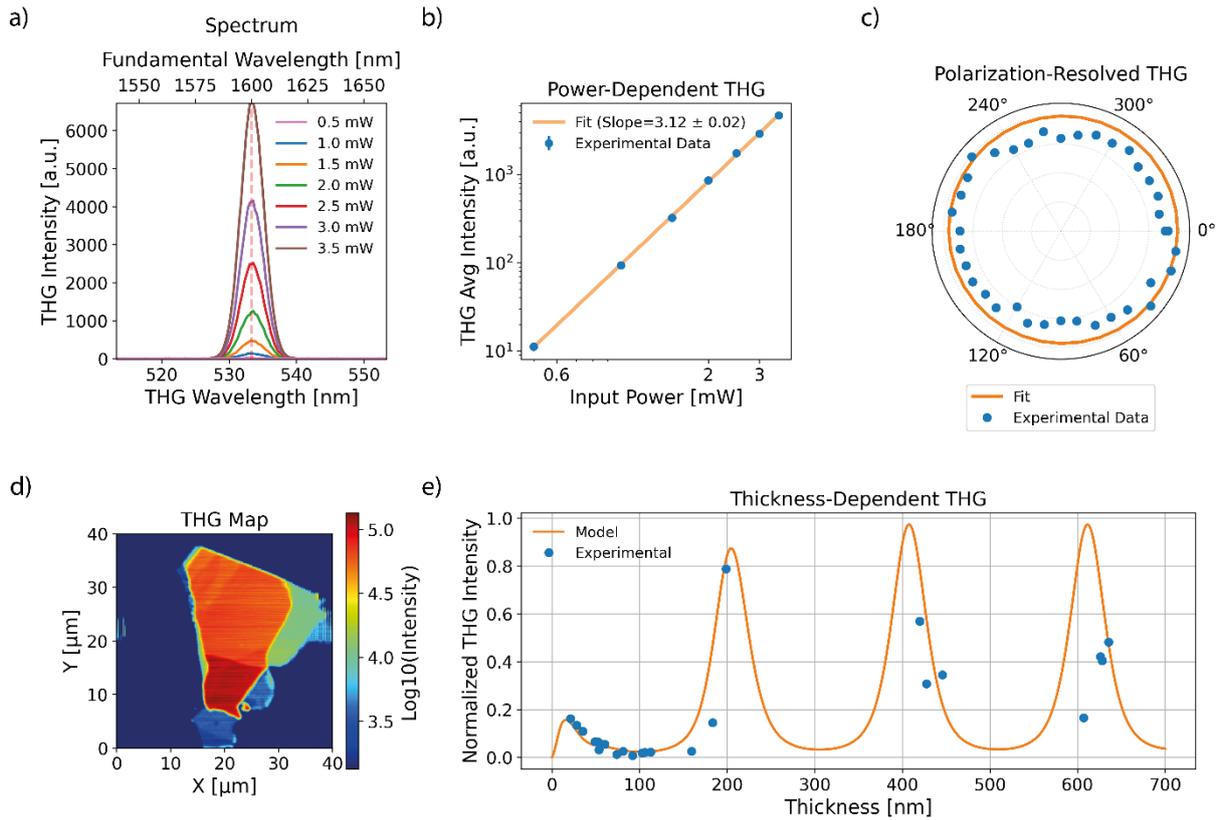

**Figure 3 Third Harmonic Generation (THG) charactrization of the 3R-MoS₂** a) Third harmonic generation (THG) spectra of 1600nm excitation wavelength for different input power. Orange dashed line shows the center of the spectrum at ~533nm. b) power dependence of THG intensity, plotted on a log-log scale. The THG intensity exhibits a power dependence with a fitted exponent of 3.12 which is in close agreement with the theoretical value of 3 predicted by electric dipole theory. c) polarization resolved THG d) THG microscopic measurement. e) thickness dependence THG of the 3R-MoS₂ at 1600nm fundamental wavelength. Theoretical model of THG efficiency compared to the experimentally measured THG efficiency of the 3R-MoS₂ sample used in this study.

**Figure 3**e demonstrates the THG thickness dependence for the same 3R-MoS₂ flakes. The thickness dependent THG intensity was simulated using the 2D wave-optics module of the COMSOL finite element method solver. To simulate the response of a plane incident on a 3R-MoS₂ thin film, we used Floquet boundary conditions in the transversal direction and perfectly matched layer domains in the longitudinal direction. For every thickness value of the thin film, we first performed a study at the fundamental wavelength to obtain the nonlinear polarization inside the 3R-MoS₂ film at the TH wavelength. This was then inserted into a study at the TH wavelength as a polarization domain. To obtain the intensity values at the harmonic wavelength, the solution fields were exported and evaluated in front of the PML domains. The TH intensity is modulated as a function of the layer thickness, and the peak intensity decreases because of the dominance of the propagation loss compared to the contribution from the layer thickness.

Like SHG, THG depends on phase-matching, medium nonlinearity, and fundamental wave transmission, yielding a series of discrete resonance peaks. However, unlike SHG, we do not observe a series of resonances of increasing power and rather exhibit fixed peak power. TH wave generated from each 3R-$MoS_2$ layer has a wavelength of 533 nm and is thus well above the material's band-gap, exhibiting substantial absorption. Hence, there is only a limited length in which TH can coherently built up, limiting the total TH power which can be generated.

## 3. Conclusions

In this work, we have demonstrated that thickness determination based on linear optical spectra provides a highly precise and scalable approach for characterizing members of the transition metal dichalcogenide (TMD) family like 3R-$MoS_2$. This method enables the accurate quantification of layer thickness across larger areas, and on substrates such as PDMS which is of central importance for the integration of verified exfoliated layers in optical systems, with an average deviation of less than two nanometers.

Building on this characterization, we investigated the nonlinear optical response of 3R-$MoS_2$ and observed the presence of a strong thickness dependence of second-harmonic generation (SHG) and third-harmonic generation (THG). The simultaneous appearance of the thickness-modulated SHG and THG in this material can be attributed to its high refractive index, which gives rise to pronounced Fabry-Pérot interference effects within the layered structure. These interference effects substantially modify the local optical field distributions and thereby strongly influences the efficiency of harmonic generation processes.

Our results show that accurate sample thickness control is essential for using TMDs in nonlinear optics and quantum photonics. We show that unless one compensates for thickness-dependent interference effects by e.g. introducing highly efficient anti-reflection coatings, the nonlinear conversion efficiency remains strongly governed by the optical path length and corresponding resonance conditions. While this general dependency applies to both SHG and THG, our analysis shows that increasing thickness beyond the first resonance is only beneficial when the generated harmonic wavelength lies below the material's absorption band edge, as absorption otherwise counteracts the constructive interference that would enhance the nonlinear signal.

Taken together, these results establish thickness as a key design parameter for optimizing the nonlinear optical response of layered TMD materials. The demonstrated correlation between thickness, linear spectra, and harmonic generation efficiency provides a powerful framework for tailoring device performance. Beyond its relevance for fundamental studies of light-matter interactions in van der Waals materials, our work highlights the central role of precise thickness control in advancing nonlinear and quantum applications, including frequency conversion, quantum light sources, and integrated photonic platforms based on TMDs.

## 4. Material and Methods

*Sample preparation:* 3R-MoS$_2$ flakes were exfoliated mechanically from bulk 3R-MoS$_2$ (hq-graphene) onto a PDMS substrate using the well-established scotch-tape method [43]. A small amount of 3R-MoS$_2$ was first placed on a piece of scotch tape, then the tape was folded and unfolded repeatedly to thin down the layers. Then the tape was carefully pressed onto a pre-cleaned PDMS substrate to successfully transfer the 3R-MoS$_2$ flakes. The process requires careful handling to maintain the uniform flakes. After transferring, the desired flakes were chosen under a brightfield optical microscope.

*Reflectivity Measurements:* The reflectivity spectra were acquired using a Zeiss microscope coupled to a Horiba spectrometer and a cooled Si CCD detector, using a broadband Xenon short-arc light source covering 240-2400nm to illuminate the sample through an 50X objective. An adjustable pinhole was used to confine the region of interest. All optical and acquisition settings were held constant through the study. The spectra were recorded through the setup's reliable operating window 500-950nm and subsequently normalized to the protected silver mirror reference to yield absolute reflectivity. To confirm both reproducibility and spatial uniformity, measurements were repeated many times, although for clarity we present here a single representative dataset for each position.

*AFM measurements:* Atomic Force Microscopy (AFM) measurements were conducted utilizing a Dimension Edge system (Bruker) operated in tapping mode to minimize sample perturbation and enhance imaging fidelity. High-resolution scans were acquired using Tap300Al-G silicon cantilevers, optimized for tapping mode, with nominal resonance frequencies (300Hz) and force constants suitable for topographic characterization. The scanning area was set to $20 \times 20$ μm$^2$, providing a broad field of view, while image acquisition was performed at a resolution of $512 \times 512$ pixels to ensure detailed surface morphology analysis. A scan rate of 0.5 Hz was employed to balance imaging speed with spatial resolution and noise suppression.

*Bland-Altman analysis:* Agreement between AFM and reflectometry was assessed using Bland–Altman analysis. For each co-located region of interest, the thickness difference $\Delta i = t_i(R) - t_i(A)$ was calculated. The mean difference $\Delta$ represents the bias between the two methods. The **limits of agreement (LoA)** were defined as $\Delta \pm 1.96, s\Delta$ where $s\Delta$ is the standard deviation of the differences, indicating the range within which 95% of the method-to-method differences are expected to lie.

*Nonlinear measurement:* The SHG and THG measurements were performed using a custom-built nonlinear microscope setup. The fundamental excitation beam was provided by a Lightconversion Carbide femtosecond laser (Model: CB3-80W) delivering pulses of 211 fs duration at a repetition rate of 2000 kHz, with a tunable wavelength range of 315- 2600nm through OPA (Lightconversion Orpheus). The excitation wavelength was fixed at 1600 nm for this measurement. as shown in **Figure 1**a, the laser beam was directed first through a half-wave plate (HWP) mounted on a rotation stage and fixed linear polarizer to control the input power, then through another HWP mounted on a rotation stage to control the incident polarization. The beam was then focused onto the sample using a long working distance Mitutoyo Plan Apochromat objective lens (0.42 NA, 50X magnification). The sample was mounted on a XYZ PI-piezoelectric stage to ensure precise spatial alignment for the mapping. The SHG and THG signal generated in transmission was collected by the same model, long working distance Mitutoyo Plan Apochromat objective lens (0.42 NA, 50X magnification) and after passing through appropriate filters (detection filters for SHG: short pass filter @950nm, short pass filter @850nm, long pass filter @700nm - detection filters for THG : 2 short pass filters @600nm- 2 band pass filters 335-610nm and 2 short pass filters @850nm) to suppress the fundamental laser light, directed to a sCMOS Andor camera (Model: Zyla 4.2),

For polarization-resolved SHG and THG, the polarization state of the detected SHG and THG signal was analysed using a linear polarizer mounted on the rotation stage. The SHG intensity map was recorded while the input polarization and analyser were fixed along the armchair direction of each flake. The power of the excitation beam at the sample was maintained at 1mW before objective.

# Supporting Information:

**Thickness-dependence of linear and nonlinear optical properties of multilayer 3R-MoS$_2$**

*Fatemeh Abtahi[1], Alen Shaji[1], Gia Quyet Ngo[1,2], Benjamin Laudert[1], Hossein Esfandiar[2], Sebastian W. Schmitt[1,2], and Falk Eilenberger[1,2,3]*

[1] *Institute of Applied Physics, Abbe Center of Photonics, Friedrich Schiller University Jena, Albert-Einstein-Str. 15, 07745 Jena, Germany*

[2] *Fraunhofer Institute for Applied Optics and Precision Engineering IOF*

*Albert-Einstein-Str. 7, 07745 Jena, Germany*

[3] *Max Planck School of Photonics, Albert-Einstein-Str. 15, 07745 Jena, Germany*

**S1. 3R-MoS$_2$ thickness calculation through reflectivity measurement**

S1.1 Reflectivity Formula

We applied Fresnel's equations together with the transfer matrix method to describe the linear optical response of the flakes. Specifically, our three- layer model, including air, 3R-MoS$_2$ and SiO$_2$, allows the calculation of the reflectivity R:

$$R = \left|\frac{(k_1 M_{22} - k_3 M_{11}) - i(M_{21} + k_1 k_3 M_{12})}{(k_1 M_{22} + k_3 M_{11}) + i(M_{21} - k_1 k_3 M_{12})}\right|^2 \qquad \text{Eq-S 1}$$

$k_i = 2\pi n_i / \lambda$ is the wavevector the related medium. Here, $k_1, k_2$ and $k_3$ represent the wavevectors of the air, 3R-MoS$_2$ and SiO$_2$, respectively. The refractive index ($n_i$) of the 3R-MoS$_2$ was taken from literature[17], and the refractive index of SiO$_2$ was taken from the Sellmeier equation. $M_{ij}$ stands for the transfer Matrix elements and describes how the light propagates through the thin film and interacts with its boundaries. The elements of the transfer matrix are as follows:

$$M_{11} = cos(k_2 d), M_{12} = \frac{1}{2}sin(k_2 d), M_{21} = -k_2 sin(k_2 d), M_{22} = cos(k_2 d) , \text{Eq-S 2}$$

d is the thickness of the 3R-MoS$_2$.

S1.2 Validation across additional thicknesses: Reflectivity vs AFM

As another example of our 3R-MoS$_2$ flakes for our reflectivity-based thickness measurement, in Figure S **1** we show a flake with several thicknesses 25 to 112nm. We marked 5 different domains of this flake and measured the reflection spectra of each domain while it was on the PDMS substrate and also after transferring on the SiO$_2$. The thickness of each domain at the end measured with the AFM method while the flake is on SiO$_2$.

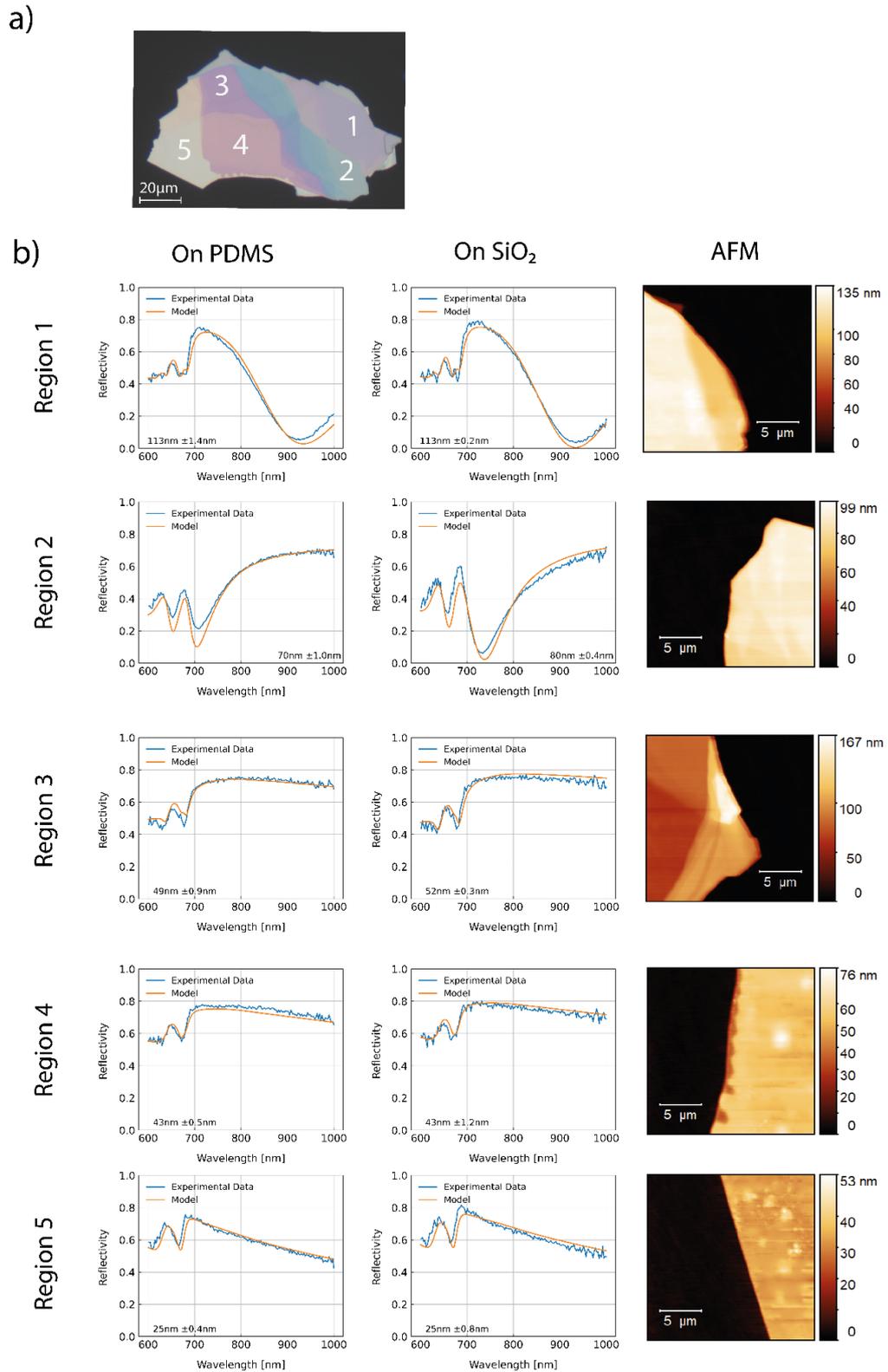

**Figure S 1 Reflectivity based thickness measurement for a sample with several thicknesses.** a) microscopic image of the 3R-MoS$_2$ flake on PDMS substrate with 5 different marked domains. b) the reflectivity spectra of each of marked domains on PDMS and SiO$_2$ substrate. blue lines are for the experimental reflection spectra of each domain, and the orange lines are the fitted model to each reflectivity spectra to obtain the desire domain's thickness. And last columns is AFM map contains of each domain to confirm our reflectivity results.

S1.2 Additional dataset for method comparison

In this section, we also present extended reflectivity results that complement **Figure 1** in the main text.

**Table S 1** thickness measured via reflectivity vs AFM

| Domains | Thickness on PDMS [nm] | ΔT of fitting on PDMS [nm] | Thickness on SiO$_2$ [nm] | ΔT of fitting on SiO2 [nm] | AFM Thickness [nm] | ΔT = RMS$_{h_1}$+RMS$_{h_1}$[nm] |
|---|---|---|---|---|---|---|
| 1 | 3 | 0.2 | 3 | 0.3 | 3 | 1 |
| 2 | 19 | 1.1 | 19 | 0.2 | 20 | 2 |
| 3 | 25 | 0.4 | 25 | 0.8 | 31 | 4.2 |
| 4 | 44 | 0.5 | 43 | 1.2 | 47 | 1.4 |
| 5 | 49 | 0.9 | 52 | 0.3 | 53 | 1.6 |
| 6 | 64 | 0.8 | 64 | 0.5 | 60 | 4.6 |
| 7 | 74 | 0.4 | 76 | 0.2 | 75 | 2.9 |
| 8 | 70 | 1 | 80 | 0.4 | 80 | 1.8 |
| 9 | 92.0 | 0.2 | 92.0 | 0.2 | 92.0 | 3.7 |
| 10 | 104 | 0.5 | 104 | 1 | 103 | 3.7 |
| 11 | 113 | 1.4 | 113 | 0.2 | 112 | 6.5 |
| 12 | 161 | 0.2 | 167 | 0.1 | 166 | 1.4 |
| 13 | 185 | 1 | 185 | 0.8 | 185 | 2.4 |
| 14 | 194 | 1 | 194 | 0.4 | 195 | 5.9 |
| 15 | 200 | 0.8 | 200 | 1 | 200 | 2.2 |

**S1.3 relative error and Bland-Altman calculation**

Agreement between Reflectivity and AFM method was evaluated with Bland-Altman analysis (Figure S **2**-a). This method used to determine the offset of reflectivity-based thickness measurements from AFM result which reported in the main text.

The relative error for each method is shown in Figure S **2**-b. AFM exhibited the relative error from 0.8 to 33.3%, with the highest error observed at thickness of 3nm. Reflectivity method exhibit relative errors of 0.1-6.7%. For thicknesses above 30nm, reflectivity showed substantially higher precision, with relative errors consistently below 10%, while AFM relative errors are still higher.

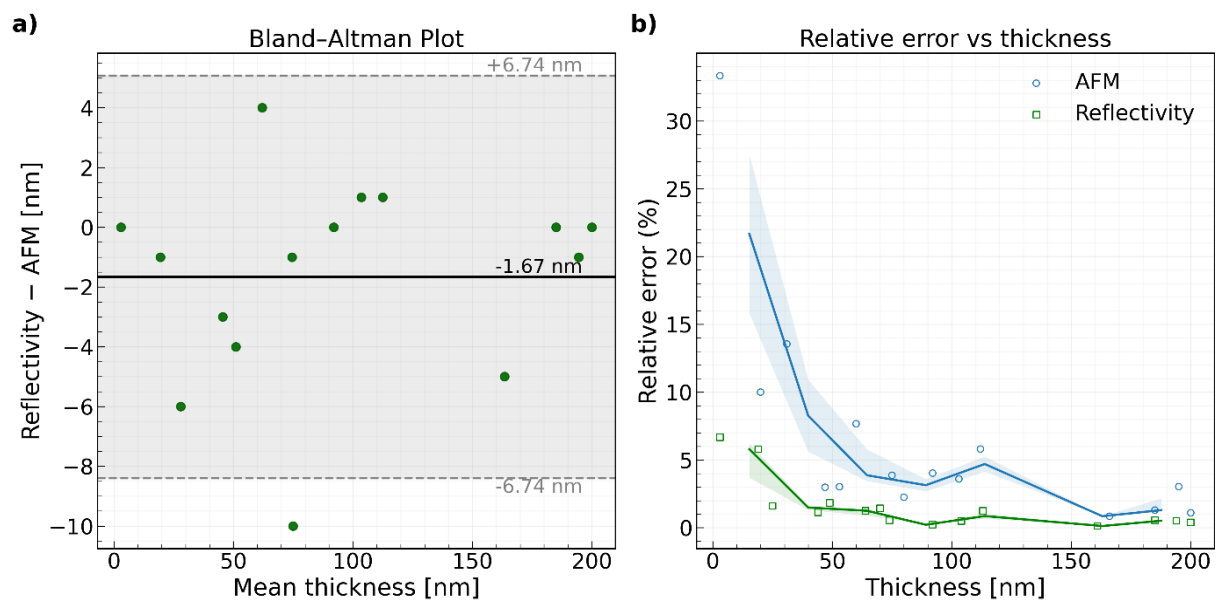

**Figure S 2 Comparison of Reflectivity results to AFM**. a) Bland-Altman plot shows a mean difference of -1.67nm with 95% limit of agreement [-8.40, 5.07] nm. b) relative error range of AFM (from 0.8-33.3%) and Reflectivity (from 0.1-6.7%).

**S2. Wavelength dependence spectrum for SHG and THG**

Before starting our nonlinear measurements, we did the Fundamental wavelength sweep measurement from 1530nm to 165nm for one sample and measured both second and third harmonic generation. According to the results of this measurement, we chose 1600nm as our fundamental wavelength which results to higher third harmonic generation intensity.

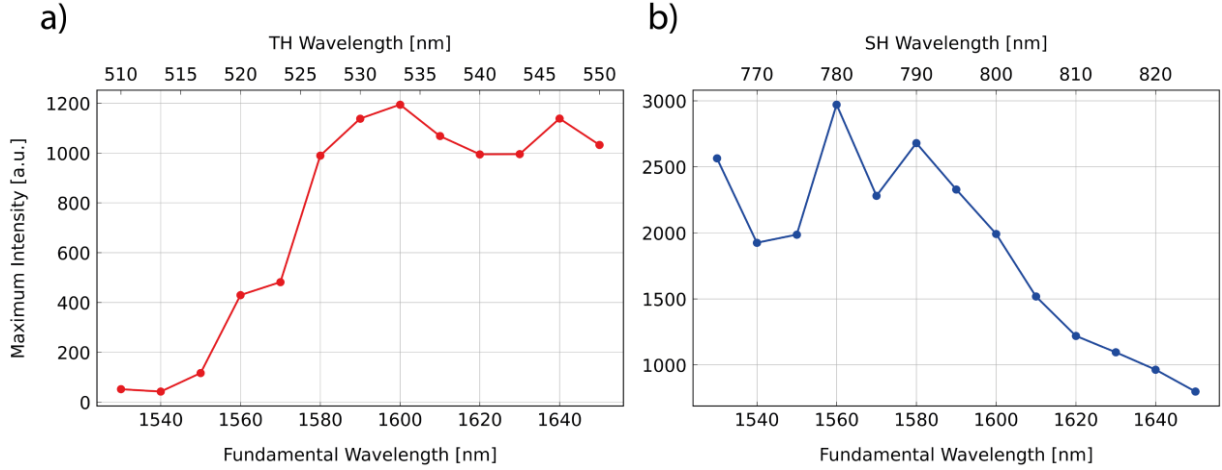

**Figure S 3 the maximum intensity** of a) the THG (red) and b) SHG (blue) spectrum is plotted for excitation wavelength range 1530nm-1650nm with step of 10nm- shows the maximum THG for the example 3R-MoS$_2$ flake at 1600nm and the SHG peak is at 1560nm.

**S3. SHG model**

We have modeled the dependence of SHG on the thickness of 3R-MoS$_2$ assuming that the three-layer system (SiO$_2$/MoS$_2$/air) with refractive indices n$_0$/n$_1$/n$_2$ forming a cavity which will modulate the fundamental wave and hence, influence the behaviour of the SH wave. The experiment was performed in transmission. Accordingly, the transmissivity of FW light can be expressed as:

$$T_\omega = \frac{Re\{n_2\}}{Re\{n_0\}} \left| \frac{t_{01}t_{12}}{e^{ik_1 h} + r_{01}r_{12}e^{-ik_1 h}} \right|^2 \quad \text{Eq-S 3}$$

Where t$_{ij}$ and r$_{ij}$ are the transmissivity and reflectivity coefficients from layer i to layer j, respectively; k$_1$ is the wavevector at the MoS$_2$ layer; and h is the thickness of MoS$_2$.

A nonlinear process like SHG is heavily dependent on the phase-matching condition, the nonlinearity of the medium and the amplitude of the excitation field. The dispersion of the MoS$_2$ layer together with the increase of the thickness will introduce a phase mismatch, which can be calculated as[17]

$$I_{2\omega} \propto |\chi^{(2)}|^2 T_\omega^2 I_\omega^2 h^2 \text{sinc}^2(\Delta k h/2) \quad \text{Eq-S 4}$$

Where $\Delta k = k_{2\omega} - 2k_\omega$ is the wavevector mismatch between the FW and SH and $\chi^{(2)}$ is the second-order susceptibility of MoS$_2$. Since $\chi^{(2)}$ is constant at a fixed FW and the excited power of fundamental wave is unchanged, the modulation of the SHG is merely dependent on the modulation of T$_\omega$ and the thickness h.

In the experiment with SHG, the second harmonic intensity modulates as a function of the 3R-MoS$_2$ thickness, and the peak intensity increases because of the dominance of thickness h, as shown in **Error! Reference source not found.**, while the propagation loss at SH is almost negligible because the SH at 800 nm has a lower energy than the band gap of $MoS_2$.

**S4. THG model**

The thickness dependent THG intensity was simulated using the 2D wave-optics module of the COMSOL finite element method solver. To simulate the response of a plane incident on a 3R-$MoS_2$ thin film, we used Floquet boundary conditions in the transversal direction and perfectly matched layer domains in the longitudinal direction. For every thickness value of the thin film, we first performed a study at the fundamental wavelength to obtain the nonlinear polarization inside the 3R-$MoS_2$ film at the third harmonic (TH) wavelength. This was then inserted into a study at the TH wavelength as a polarization domain. To obtain the intensity values at the harmonic wavelength, the solution fields were exported and evaluated in front of the PML domains. The TH intensity is modulated as a function of the layer thickness, and the peak intensity decreases because of the dominance of the propagation loss compared to the contribution from the layer thickness. Accordingly, THG could serve as a further method for thickness characterization of 2D materials.

## S5. Polarization-resolved and map of second and third harmonic generation on more samples

As an additional results, we show here the orientation of 6 other samples which we did SHG and THG on them.

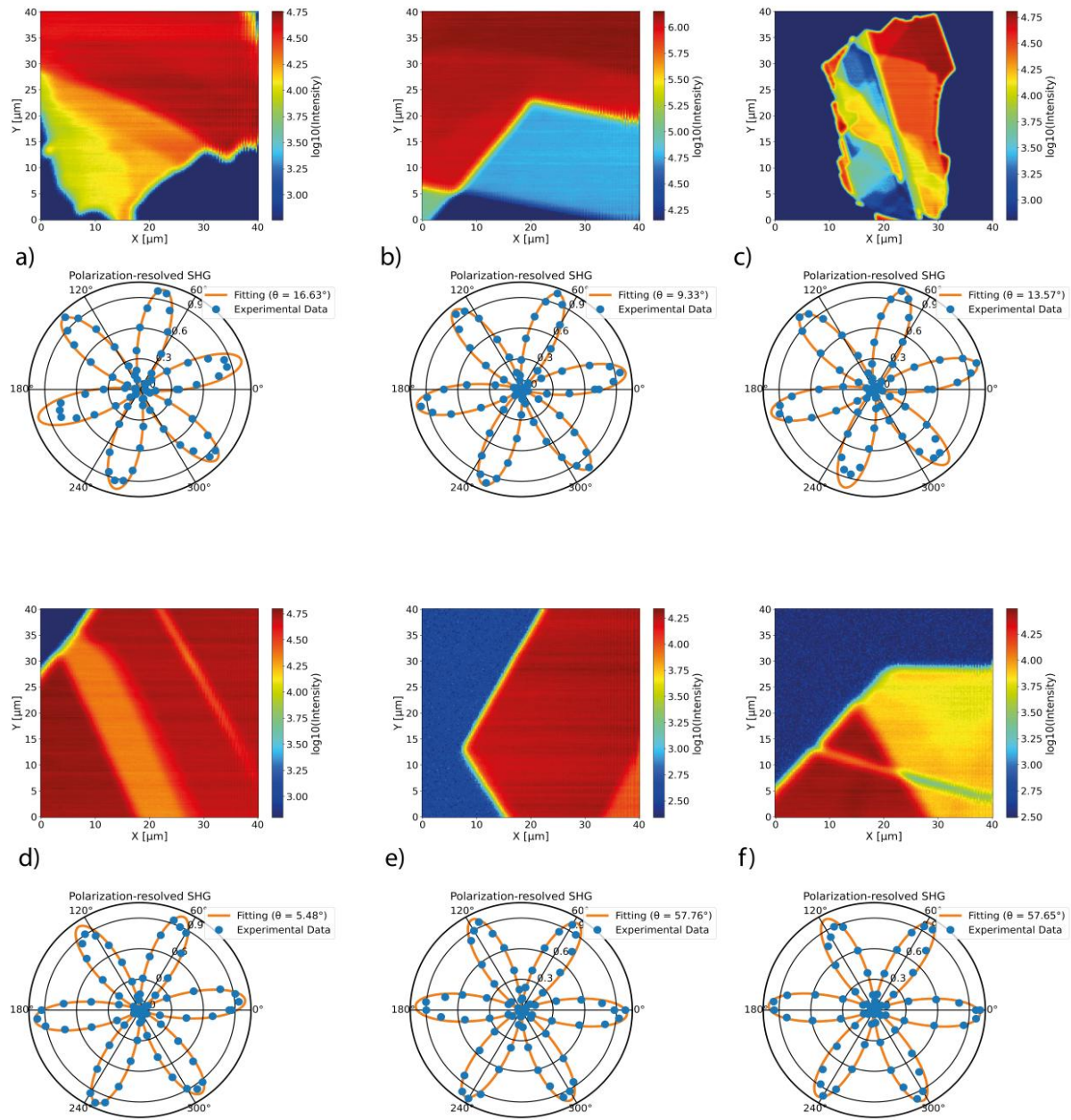

**Figure S 4 SHG map** with marked regions and polarization dependency of remain flakes which shows the orientation of each of them.

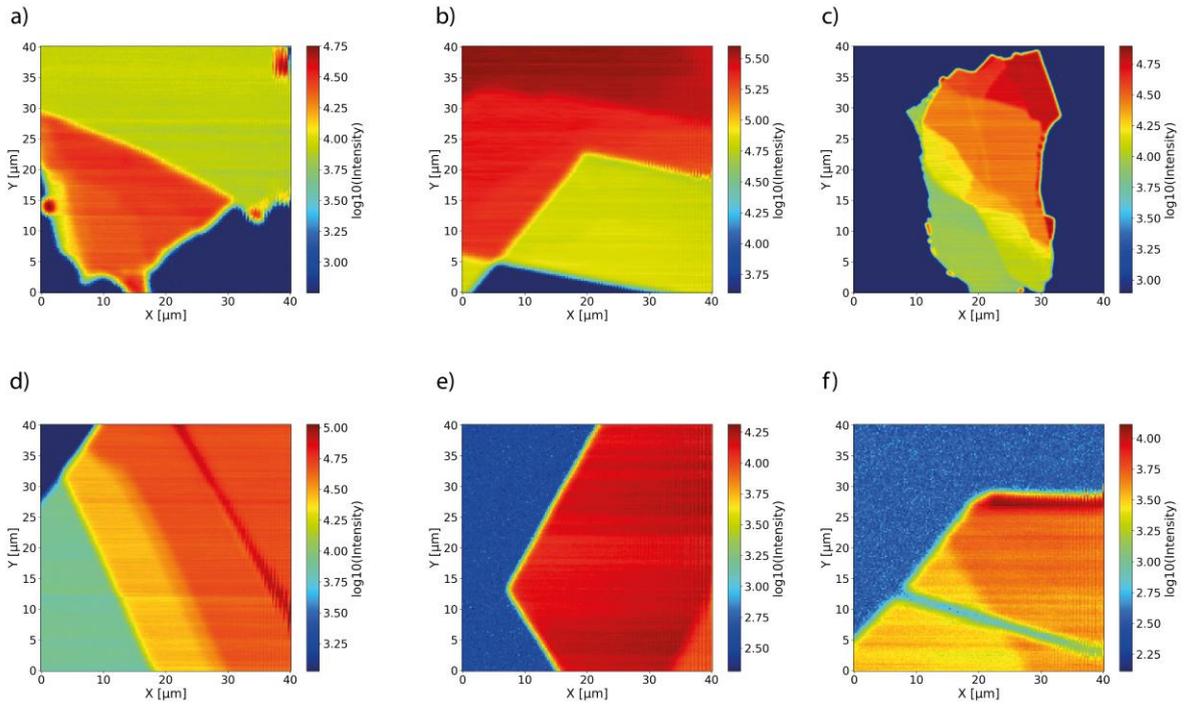

**Figure S 5 THG map of the flakes**